\documentclass[12pt]{article}
\newcommand{\RRR}{$I\hspace{-0.25em}R^3$}
\newcommand{\ket}[1]{\vert#1\rangle}

\newcommand{\braket}[2]{\langle#1\vert#2\rangle}
\newcommand{\ketbra}[2]{\vert#1\rangle\langle#2\vert}

\newcommand{\bd}{\begin{displaymath}}
\newcommand{\ed}{\end{displaymath}}
\newcommand{\be}{\begin{equation}}
\newcommand{\ee}{\end{equation}}
\newcommand{\bi}{\begin{itemize}}
\newcommand{\ei}{\end{itemize}}
\newcommand{\bq}{\begin{quote}}
\newcommand{\eq}{\end{quote}}

\begin{document}
\setlength{\baselineskip}{16pt}
\title{Is the end in sight for theoretical pseudophysics?}
\author{\\Ulrich Mohrhoff\\
Sri Aurobindo International Centre of Education\\
Pondicherry 605002 India\\
\normalsize\tt ujm@auromail.net}
\date{}
\maketitle
\begin{abstract}
\noindent The question of what ontological message (if any) is encoded in the 
formalism of contemporary physics is, to say the least, controversial. The reasons 
for this state of affairs are psychological and neurobiological. The processes by 
which the visual world is constructed by our minds, predispose us towards 
concepts of space, time, and substance that are inconsistent with the 
spatiotemporal and substantial aspects of the quantum world. In the first part of 
this chapter, the latter are extracted from the quantum formalism. The nature of a 
world that is fundamentally and irreducibly described by a probability algorithm is 
determined. The neurobiological processes responsible for the mismatch between 
our ``natural'' concepts of space, time, and substance and the corresponding 
aspects of the quantum world are discussed in the second part. These natural 
concepts give rise to pseudoproblems that foil our attempts to make ontological 
sense of quantum mechanics. If certain psychologically motivated but physically 
unwarranted assumptions are discarded (in particular our dogged insistence on 
obtruding upon the quantum world the intrinsically and completely differentiated 
spatiotemporal background of classical physics), we are in a position to see why 
our fundamental physical theory is a probability algorithm, and to solve the 
remaining interpretational problems.
\setlength{\baselineskip}{15pt}
\end{abstract}
\newpage

\hfill\parbox[b]{4in}{The most satisfying way to end a philosophical dispute is to 
find a false presupposition that underlies all the puzzles it involves.\par
\hfill B.C. Van Fraassen~\cite{VF}}

\vspace{0.5cm}\section{\large INTRODUCTION}
According to the ``minimal instrumentalist interpretation''~\cite{Redhead}, the 
mathematical formalism of quantum mechanics (QM) is an algorithm for assigning 
probabilities to the possible outcomes of measurements that may be made, on the 
basis of actual measurement outcomes. Is there more to QM than that? If QM is our 
fundamental theoretical framework, then the answer must surely be affirmative. 
But if the answer is affirmative, how can our fundamental theoretical framework be 
a probability algorithm? While a ``mere'' probability algorithm does not appear 
capable of describing the events to which it assigns probabilities, a fundamental 
theory must encompass these events as well as the correlations between them.

The first aim of this chapter is to show how QM can be both our fundamental 
theoretical framework and a ``mere'' probability algorithm. The task of 
determining the nature of a world that is fundamentally and irreducibly described 
by a probability algorithm, is carried out in Secs. 2--6 and resumed in Sec.~10. 
What transpires there is in conflict with some of our deepest convictions about 
space, time, and matter. The second aim of this chapter is to determine the 
psychological and neurobiological origins of this conflict, which is responsible for 
the difficulties we face in trying to make sense of QM. This task is undertaken in 
Secs.~7 and~9.

The visual world is constructed by our minds in conformity with the Cookie Cutter 
Paradigm, which says, in effect, that the synchronic multiplicity of the world rests 
on surfaces that carve up space much as cookie cutters carve up rolled-out pastry. 
The most nefarious consequence of this biologically ``hardwired'' paradigm is our 
dogged insistence on obtruding upon the quantum world the intrinsically and 
completely differentiated spatiotemporal background of classical physics (Sec.~8). 
This, more than anything else, is responsible for the belief that the wave function 
represents an evolving instantaneous physical state, and hence for the mother of 
all pseudoproblems---how to explain (away) the discontinuous mode of evolution 
associated with measurements (Sec.~9). What obtains between two successive 
measurements is not an evolving physical state but a fuzzy state of affairs that is 
temporally undifferentiated (Sec.~10).

What remains of the so-called ``measurement problem'' (including the so-called 
``pointer problem'') when this is stripped of the notion that quantum states 
evolve, is solved in Sec.~11. Comments on three interpretative strategies 
(spontaneous localization theories, decoherence theories, and modal 
interpretations) are offered in Sec.~12. The penultimate section deals with the 
limits that QM imposes on explanations, and in the concluding section I explain 
why I am afraid that the answer to the question posed in the title of this chapter is 
negative.

\section{\large THE RELATIVE AND CONTINGENT REALITY OF\\
SPATIAL DISTINCTIONS}
\label{rcr}
Consider the probability distribution $|\psi(x)|^2$ associated with the position of 
the electron relative to the nucleus in a stationary state of atomic hydrogen. 
Imagine a small region~$V$ for which $\int_V|\psi(x)|^2dx$ differs from both 0 
and~1. While the atom is in this state, the electron is neither inside~$V$ nor 
outside~$V$. (If the electron were inside, the probability of finding it outside 
would be~0, and vice versa.) But being inside~$V$ and being outside~$V$ are the 
only possible relations between the electron and~$V$. If neither relation holds, 
this region simply does not exist for the electron. It has no reality as far as the 
electron is concerned. But conceiving of a region~$V$ is tantamount to making the 
distinction between ``inside~$V$'' and ``outside~$V$''. Hence instead of saying 
that $V$ does not exist for the electron, we may say that the distinction we make 
between ``inside~$V$'' and ``outside~$V$'' is a distinction that the electron does 
not make. Or we may say that the distinction we make between ``the electron is 
inside~$V$'' and ``the electron is outside~$V$'' is a distinction that Nature does 
not make. It corresponds to nothing in the physical world. It exists only in our 
heads.

Suppose, again, that the observables $A$, $B$, and~$C$ are consecutively 
measured, the outcome of the first measurement being~$a$. If the Hamiltonian 
is~0, the probability of finding~$c$ after the intermediate measurement of~$B$ is
\be
\label{e0}
p\,(c|a,B)=\sum_i\left|\braket c{b_i}\braket{b_i}a\right|^2.
\ee
If the Hamiltonian is not~0, the brackets are transition amplitudes. Formula 
(\ref{e0}) applies whenever information concerning the value of~$B$ is in 
principle available, however difficult it may be to obtain it. If such information is 
strictly unavailable, the probability of finding $c$ is
\be
p\,(c|a)=\left|\braket ca\right|^2=\left|\sum_i\braket 
c{b_i}\braket{b_i}a\right|^2.
\ee
Thus in one case we first calculate the absolute squares of the amplitudes $\braket 
c{b_i}\braket{b_i}a$ and then add the results, and in the other case we first add 
the amplitudes associated with the alternatives (which are defined in terms of 
possible measurement outcomes) and then square the absolute value of the result. 
Why? Because in one case the distinctions we make between the alternatives have 
counterparts in the physical world, and in the other case they don't. To cite a 
familiar example, in one case the electron~($e$) goes through either the left 
slit~($L$) or the right slit~($R$). The propositions ``$e$~went through~$L$'' and 
``$e$~went through~$R$'' possess truth values: one is true, the other is false. In 
the other case, these propositions lack truth values; they are neither true nor false 
but meaningless. All we can say in this case is that it goes through $L\&R$, the 
region defined by the slits considered as a whole. We cannot say more than that 
because the distinction we make between ``$e$~went through~$L$'' and 
``$e$~went through~$R$'' is a distinction that Nature does not make in this case. 
It has no counterpart in the actual world.

Thus whenever QM requires us to add amplitudes, the distinctions we make 
between the corresponding alternatives are distinctions that correspond to nothing 
in the physical world. This implies, in particular, that the reality of the spatial 
distinctions we make is relative and contingent. ``Relative'' because our 
distinction between the inside and the outside of a region may be real for a given 
object at a given time, and it may have no reality for a different object at the same 
time or for the same object at a different time; and ``contingent'' because the 
existence of a given region~$V$ for a given object~$O$ at a given time~$t$ 
depends on whether the proposition ``$O$~is in~$V$ at the time~$t$'' has a truth 
value.

\section{\large THE IMPORTANCE OF THE ``MEASUREMENT\\
APPARATUS''}
\label{impdet}
Does this proposition have a truth value if none is indicated (by an actual event or 
state of affairs)? To find out, suppose that $W$ is a region disjoint from~$V$, and 
that $O$'s presence in~$V$ is indicated. Isn't $O$'s absence from~$W$ indicated 
at the same time? Are we not entitled to infer that the proposition ``$O$~is 
in~$W$'' has a truth value (namely, ``false'')? Because the reality of spatial 
distinctions is relative and contingent, the answer is negative. Regions of space do 
not exist ``by themselves''. The distinction we make between ``inside~W'' and 
``outside~W'' has no physical reality {\it per se\/}. If $W$ is not realized (made 
real) by some means, it does not exist. But if it does not exist, the proposition 
``$O$~is in~$W$'' cannot have a truth value. All we can infer from $O$'s indicated 
presence in~$V$ is the truth of a {\it counterfactual\/}: if~$W$ were the 
sensitive region of a detector~$D$, $O$~would not be detected by~$D$. 
Probability~1 is not sufficient for ``is'' or ``has''.

It follows that a detector%
\footnote{A perfect detector, to be precise. If $D$ is less than 100~percent 
efficient, the absence of a click does not warrant the falsity of ``$O$~is in~$W$''.}
performs two necessary functions: it indicates the truth value of a proposition of 
the form ``$O$~is in~$W$'', and by realizing~$W$ (or the distinction between 
``inside~$W$'' and ``outside~$W$'') it makes the predicates ``inside~$W$'' and 
``outside~$W$'' available for attribution to~$O$. The apparatus that is 
presupposed by every quantum-mechanical probability assignment is needed not 
only for the purpose of indicating the possession, by a material object, of a 
particular property (or the possession, by an observable, of a particular value) but 
also for the purpose of realizing a set of properties or values, which thereby 
become available for attribution. This does not mean that QM~is restricted ``to be 
exclusively about piddling laboratory operations''~\cite{BellAM}. Any event from 
which either the truth or the falsity of a proposition of the form ``system~$S$ has 
property~$P$'' can be inferred, qualifies as a measurement.

\section{\large THE INCOMPLETE SPATIOTEMPORAL\\
DIFFERENTIATION OF THE PHYSICAL WORLD}
\label{finidiff}
Let \RRR($O$) be the set of (purely imaginary) exact positions relative to an 
object~$O$. Since no material object ever has a sharp position, we can conceive of 
a partition of \RRR($O$) into finite regions that are so small that none of them is 
the sensitive region of an actually existing detector. Hence we can conceive of a 
partition of \RRR($O$) into sufficiently small but finite regions~$V_i$ of which the 
following is true: there is no object~$Q$ and no region~$V_i$ such that the 
proposition ``$Q$~is inside~$V_i$'' has a truth value. In other words, there is no 
object~$Q$ and no region~$V_i$ such that $V_i$ exists for~$Q$. But a region of 
space that does not exist for any material object, does not exist at all. The 
regions~$V_i$ represent spatial distinctions that Nature does not make. They 
correspond to nothing in the physical world. It follows that the world is not 
infinitely differentiated spacewise. Its spatial differentiation is incomplete---it 
doesn't go all the way down.

While positions are indicated by (macroscopic) detectors, times are indicated by 
(macroscopic) clocks. (The word ``macroscopic'' is defined in Sec.~\ref{profac}.) 
Since clocks indicate times by the positions of their hands,%
\footnote{Digital clocks indicate times by transitions from one reading to another, 
without hands. The uncertainty principle for energy and time implies that such a 
transition cannot occur at an exact time, except in the unphysical limit of infinite 
mean energy~\cite{Hilge}.}
and since sharp positions do not exist, neither do sharp times. From this the 
incomplete temporal differentiation of the physical world follows in exactly the 
same way as its incomplete spatial differentiation follows from the nonexistence of 
sharp positions. Neither the spatial nor the temporal differentiation of the world 
goes all the way down.

\section{\large SPACE AND MATTER: SELF-RELATIONS AND\\
NUMERICALLY IDENTICAL RELATA}
\label{sam}
If two indistinguishable particles scatter elastically, the question of which incoming 
particle is identical with which outgoing particle, cannot be answered, as is well 
known. If we label the incoming particles $A$ and~$B$ and the outgoing ones $C$ 
and~$D$, there are two alternatives: $(A{\rightarrow}C, B{\rightarrow}D)$ and 
$(A{\rightarrow}D, B{\rightarrow}C)$. Since QM requires us to add their 
amplitudes, the distinction we make between them is a distinction that Nature does 
not make. The distinction between {\it this\/} particle and {\it that\/} particle 
(over and above the distinction between {\it this\/} property and {\it that\/} 
property) corresponds to nothing in the physical world.

For centuries philosophers have argued over the existence of distinct substances. 
QM has settled the question for good: there is only one substance; it is illegitimate 
to interpose a multitude of distinct substances between the substance that 
betokens existence and the multitude of existing (``possessed'') positions.%
\footnote{Even such a seemingly harmless expression as ``possessed position'' is 
seriously misleading, inasmuch as it suggests the existence of a multitude of 
position-possessing substances. There is only the substance that warrants the 
existence of positions, and this is the same for all existing (``possessed'') 
positions.}
Distinctness is strictly a matter of properties. $A$,~$B$, $C$, and~$D$ label sets 
of properties, not substances. As a set of properties, each is distinct from the 
others. As substances, all are identical in the strong sense of numerical identity.

We thus arrive at the following conclusions: Space is not an intrinsically 
differentiated, pre-existent expanse, and matter is neither a collection of 
substantial fields nor a multitude of distinct substances. Instead, space is the 
totality of existing spatial relations%
\footnote{Hence there is no such thing as empty space---not because space is 
``filled with vacuum fluctuations'', as some say, but because there are no 
unoccupied locations or unpossessed positions. Where there is nothing (no thing) 
there is no ``there''.}
(``possessed'' relative positions and relative orientations), and matter is the 
corresponding {\it apparent\/} multitude of relata---apparent because the 
relations are {\it self\/}-relations.%
\footnote{It has been claimed that relationism---the doctrine that space and time 
are a family of spatial and temporal relations holding among the material 
constituents of the universe---is untenable on account of its failure to 
accommodate inertial effects. See Refs.~\cite{Dieks1,Dieks2} for a refutation of 
this claim.}

But are not the relations embedded in a continuous, 3-dimensional expanse? And is 
not space this continuous expanse, rather than a set of relations? Here are two 
possible answers: (i)~Spatial extension is a property of each spatial relation. What 
accounts for the spatial character of spatial relations is this property, rather than a 
pre-existent spatial expanse. If (in our minds) we abolish matter (the relata), we 
abolish space (the relations) as well, and what then remains is not a continuous 
expanse undifferentiated by relations but nothing (at any rate, nothing that is 
differentiated or extended). (ii)~If we insist on thinking of spatial extension as the 
property of a substantial expanse, we must think of this expanse as intrinsically 
devoid of multiplicity. The divisions we tend to project into it correspond to nothing 
in the physical world. Since being extended and being undivided is a rather 
paradoxical combination of properties for a substance, I prefer the first answer.

\section{\large THE TOP-DOWN STRUCTURE OF THE\\
PHYSICAL WORLD}
\label{tds}
If all that space contains (in the proper, set-theoretic sense of ``containment'') is 
spatial relations, then the form of a material object is the set of its internal spatial 
relations (the relative positions and orientations of its constituents), and a particle 
without internal structure (no constituents, no internal spatial relations) is a 
formless object. Particles without internal structure are often said to be pointlike 
instead. There are several reasons why this should not be taken to mean more than 
that fundamental particles lack internal structure.

To begin with, nothing in the formalism of QM refers to the {\it shape\/} of an 
object that lacks internal structure, and empirical data cannot possibly do so. 
(While experiments may produce evidence that a particle has structure, they 
cannot produce evidence that a particle lacks structure. {\it A fortiori\/} they 
cannot produce evidence that a particle has a pointlike form.) The notion that an 
object without internal structure has a form (which would have to be pointlike) is 
therefore unwarranted on both theoretical and experimental grounds. In addition, 
it explains nothing. In particular, it does not explain why a composite object---be it 
a nucleon, a molecule, or a galaxy---has the shape that it does, for all empirically 
accessible forms are fully accounted for by the relative positions of their material 
constituents. Instead of contributing something to our understanding of empirically 
accessible forms, the notion that a structureless particle is a pointlike object 
encumbers us with a form that is (i)~entirely unverifiable, (ii)~explanatorily 
completely useless, and (iii)~absolutely different from all empirically accessible 
forms, which are sets of spatial relations. All we can possibly gain from postulating 
that a structureless particle has a (pointlike) form, is the assurance that quarks 
and such can be visualized (if we allow that a point can be visualized). What good 
does that do, considering that atoms (the smallest things ``made of'' structureless 
quarks and structureless electrons) can {\it not\/} be visualized as they are? It is 
to the bigger things, starting with molecules, that we can attribute something like a 
form that can be visualized as it is, not to the smaller things that make up atoms.

According to Bertrand Russell~\cite{Russell}, ``substance'' is a metaphysical 
mistake, due to transference to the world-structure of the structure of sentences 
composed of a subject and a predicate. If we let QM have its say, substance is 
anything but a mistake. There is something that constitutes matter and space but 
remains undivided by its space-constituting self-relations and the corresponding 
multitude of matter-constituting relata. Since this is the only ``thing'' that exists 
independently of anything else, ``substance'' is the appropriate word, and even 
capitalization is warranted.

What is a mistake is the attempt to build reality ``from the bottom up'', on an 
intrinsically and infinitely differentiated space or spacetime, out of locally 
instantiated physical properties, or else by aggregation, out of a multitude of 
distinct substances. Reality is built ``from the top down'', by a self-differentiation 
of one Substance. By the simple device of entering into spatial relations with itself, 
this creates both space (the existing spatial relations) and matter (the 
corresponding relata). The reason why we cannot built reality from the bottom up 
is that this top-down differentiation does not ``bottom out''. If we go on dividing a 
material object, its ``constituents'' lose their individuality, and if we conceptually 
partition the physical world into sufficiently small yet finite regions, we reach a 
point where the distinctions we make between these regions no longer correspond 
to anything in the physical world. Our spatial and substantial distinctions are 
warranted by property-indicating events, and these do not license an absolute and 
unlimited objectification of spatial distinctions.

\section{\large A TALE OF TWO WORLDS}
\label{ttw}
Let us now examine the neurobiological reasons why these ontological implications 
of QM go counter to some of our deepest convictions concerning space, time, and 
matter.

It is safe to say that the following idea appears self-evident to anyone uninitiated 
into the mysteries of the quantum world: the parts of a material object (including 
the material world as a whole) are defined by the parts of the space it ``occupies'', 
and the parts of space are defined by delimiting surfaces. Because it says, in effect, 
that the synchronic multiplicity of the world rests on surfaces that carve up space 
much as cookie cutters carve up rolled-out pastry, I have dubbed this idea ``the 
Cookie Cutter Paradigm'' (CCP)~\cite{qmccp, bccp}.

The CCP is ``hard-wired'': the way in which the brain processes visual information 
guarantees that the result---the visual aspect of the phenomenal world---is a world 
of objects whose shapes are bounding surfaces. Vision is now widely recognized as 
a process of reconstruction: From optical images at the eyes, human vision 
reconstructs those properties of the physical world that are useful to the 
viewer~\cite{Marr}. The constructions of vision are based on a neural analysis of 
the visual field (the optical images falling on the retinas in both eyes) that 
capitalizes on contrast information. Data arriving from homogeneously colored and 
evenly lit regions of the visual field do not make it into conscious awareness. The 
corresponding regions of the phenomenal world are {\it filled in\/} (much like 
children's coloring books) on the basis of {\it contrast information\/} that is 
derived from {\it boundaries\/} in the visual field.

The extraction of discontinuities from the visual field (discontinuous changes in 
color and/or brightness in both the spatial and the temporal domain) begins in the 
retina itself~\cite{RoskaWerblin}. Further downstream, in the primary visual 
cortex (V1), most cells are orientation selective (in the macaque monkey about 70 
to 80 percent); the rest have center-surround receptive fields and are color 
selective%
\footnote{Orientation selective neurons respond best to lines of a particular 
orientation---bright lines on a dark background, dark lines on a bright background, 
or boundaries between areas of different color and/or brightness. The receptive 
field of a neuron is the retinal area from which input is received. A center-surround 
receptive field is divided into a small central region and a larger surrounding 
region; some neurons with such receptive fields are excited (their firing rate is 
increased) by illumination of the center and inhibited (their firing rate is 
decreased) by illumination of the surround; others are inhibited from the center 
and excited from the surround.}~\cite{Hubel}.
10~to 20 percent of the orientation selective cells are end-stopped, responding to 
short but not long line or edge stimuli. In visual area~2 (V2), at least half of the 
orientation selective cells are end-stopped~\cite{Hubel, LiHu}. Since all visual 
information flows through these cortical regions, it is eminently plausible that the 
construction of the visual world is based on line segments and involves a two-step 
synthesis: first line segments are integrated into outlines, then outlines are 
supplemented with covering surfaces much like the wire frames of CAD software.

If the shapes of phenomenal objects are bounding surfaces, and if what divides the 
phenomenal world into parts is delimiting surfaces, it is only natural that our 
conceptions of the physical world should conform to the CCP---and that we should 
be perplexed by Nature's refusal to follow suit. To my mind, the formidable 
interpretational problems associated with QM are largely due to an unreflecting 
refusal to think about the physical world along other lines than those laid down by 
the CCP. Let us now look at some of the paradigm's implications and see how they 
lead us down the garden path.

In a world whose synchronic multiplicity rests on surfaces, spatial extension exists 
in advance of multiplicity, for only what is extended can be cut up by the 
three-dimensional equivalents of cookie cutters. If, in addition, the parts of 
material objects are defined by the parts of space, then the parts of space exist in 
advance of the parts of material objects. This is how we come to think of space as a 
pre-existent and intrinsically divided expanse. But if this is how we think, we 
cannot conceive of fuzzy positions. If parts are defined by geometrical boundaries, 
the relative positions of parts are as sharply defined as their boundaries, and there 
isn't anything fuzzy about the way geometrical boundaries are defined. In an 
intrinsically and completely differentiated spatial expanse, all conceivable parts 
exist in an absolute sense, and are therefore real for all material objects. This 
means that for every material object~$O$ and for every conceivable region~$V$, 
the proposition ``$O$~is in~$V$'' possesses a truth value at all times. (In the case 
of a composite object, ``$O$''~stands for the centre-of-mass.) In other words, all 
possessed positions are sharp.

There are more ways in which our neurobiology misleads us. Although we readily 
agree that red, round, or a smile cannot exist without a red or round object or 
without a smiling face, we just as readily believe that positions can exist without 
being properties of material objects. We are prepared to think of material objects 
as substances, and we are not prepared to think of the properties of material 
objects as substances---except for one: we tend to think of positions as if they 
existed by themselves. The reasons for these disparate attitudes are to be found in 
the neurobiology of perception. They have nothing to do with the quantum world, 
other than making it hard to make sense of it.

One of these reasons is the following: The visual cortex is teeming with feature 
maps. A feature map is a layer of the cerebral cortex in which cells map a particular 
phenomenal variable (such as hue, brightness, shape, motion, or texture) in such a 
way that adjacent cells generally correspond to adjacent locations in the visual 
field~\cite{Clark}. Every phenomenal variable has a separate map (and usually not 
just one but several maps at different levels within the neuroanatomical hierarchy) 
except location, which is present in all maps. If there is a green box here and a red 
ball there, ``green here'' and ``red there'' are signalled by neurons from one 
feature map, and ``boxy here'' and ``round there'' are signalled by neurons from 
another feature map. ``Here'' and ``there'' are present in both maps, and this is 
how we know that green goes with boxy and red goes with round. Position is the 
integrating factor. In the brain, and consequently in the phenomenal world, 
positions pre-exist---in the brain at the scale of neurons, in the phenomenal world 
at visually accessible scales. They exist in advance of phenomenal objects, and this 
is another reason why we tend to assume that they also exist in advance of 
physical objects, not only at the scale of neurons or at visually accessible scales, 
but also at the scales of atoms and subatomic particles. The transition from visually 
accessible scales to subatomic scales is an unwarranted extrapolation, but if one 
postulates a pre-existent spatial expanse that is intrinsically differentiated at some 
scales, then it is hard to see why it is not intrinsically differentiated at all scales.%
\footnote{The role that position plays in perception is analogous to the role that 
substance plays in conception. Among the ideas that philosophers have associated 
with the word ``substance'', the following is relevant here: while a property is that 
in the world which corresponds to the predicate in a sentence composed of a 
subject and a predicate, a substance is that in the world which corresponds to the 
subject. It objectifies the manner in which a conjunction of predicative sentences 
with the same subject term bundles predicates. While substance thus serves as the 
``conceptual glue'' that binds an object's properties, position serves as the 
``perceptual glue'' that binds an object's phenomenal features. This analogy is 
another reason why we tend to think of positions as substances.}

\begin{table}[p]
\begin{tabular}[t]{|l|l|}
\multicolumn{2}{l}{\large A TA(B)LE OF TWO WORLDS}\\
\multicolumn{2}{l}{}\\ \hline
\rule{0pt}{20pt}\parbox[t]{6.6cm}{\bf The world according to QM\bigskip}&
\rule{0pt}{20pt}\parbox[t]{6.6cm}{\bf The world according to the 
CCP\bigskip}\\ \hline
\rule{0pt}{13pt}\parbox[t]{9.1cm}{Synchronic multiplicity rests on spatial 
relations.\smallskip}&
\rule{0pt}{13pt}\parbox[t]{6.6cm}{Synchronic multiplicity rests on delimiting 
surfaces.\smallskip}\\ \hline
\rule{0pt}{13pt}\parbox[t]{9.1cm}{Space is the totality of existing spatial 
relations.\smallskip}&
\rule{0pt}{13pt}\parbox[t]{6.6cm}{Space is an intrinsically and completely 
differentiated, pre-existent expanse.\smallskip}\\ \hline
\rule{0pt}{13pt}\parbox[t]{9.1cm}{The only existing positions are the relative 
positions between material objects.\smallskip}&
\rule{0pt}{13pt}\parbox[t]{6.6cm}{Positions exist by themselves, whether or 
not they are possessed.\smallskip}\\ \hline
\rule{0pt}{13pt}\parbox[t]{9.1cm}{All existing (=possessed) positions are 
fuzzy.\smallskip}&
\rule{0pt}{13pt}\parbox[t]{6.6cm}{All existing positions are 
sharp.\smallskip}\\ \hline
\rule{0pt}{13pt}\parbox[t]{9.1cm}{The reality of spatial distinctions is relative 
and contingent.\smallskip}&
\rule{0pt}{13pt}\parbox[t]{6.6cm}{The reality of spatial distinctions is absolute; 
they are intrinsic to a pre-existent expanse.\smallskip}\\ \hline
\rule{0pt}{13pt}\parbox[t]{9.1cm}{The spatial character of spatial relations is a 
property shared by all spatial relations.\smallskip}&
\rule{0pt}{13pt}\parbox[t]{6.6cm}{The spatial character of spatial relations 
derives from a pre-existent expanse, in which the relations are 
embedded.\smallskip}\\ \hline
\rule{0pt}{13pt}\parbox[t]{9.1cm}{A structureless particle, lacking spatial 
extent, is a formless object.\smallskip}&
\rule{0pt}{13pt}\parbox[t]{6.6cm}{A structureless particle, lacking spatial 
extent, is a pointlike object.\smallskip}\\ \hline
\rule{0pt}{13pt}\parbox[t]{9.1cm}{The form of an object with spatial extent 
consists of the object's internal spatial relations.\smallskip}&
\rule{0pt}{13pt}\parbox[t]{6.6cm}{The form of an object with spatial extent is a 
surface.\smallskip}\\ \hline
\rule{0pt}{13pt}\parbox[t]{9.1cm}{Considered by themselves, the ultimate 
constituents of matter are identical in the strong sense of numerical identity. Each 
is the one Substance that constitutes every material object.\smallskip}&
\rule{0pt}{13pt}\parbox[t]{6.6cm}{The ultimate constituents of matter are 
distinct substances.\smallskip}\\ \hline
\rule{0pt}{13pt}\parbox[t]{9.1cm}{The spatiotemporal differentiation of the 
world is incomplete.\smallskip}&
\rule{0pt}{13pt}\parbox[t]{6.6cm}{The spatiotemporal differentiation of the 
world is complete.\smallskip}\\ \hline
\rule{0pt}{13pt}\parbox[t]{9.1cm}{Owing to its incomplete spatiotemporal and 
substantial differentiation, the world has a top-down structure.\smallskip}&
\rule{0pt}{13pt}\parbox[t]{6.6cm}{The world is built from the bottom up, on an 
intrinsically and completely differentiated manifold or out of a multitude of distinct 
substances.\smallskip}\\ \hline
\end{tabular}
\end{table}
The accompanying table contrasts the salient features of the quantum world with 
the corresponding features of a world that is constructed along the lines laid down 
by the CCP. Not every one of the features in the right column can be directly linked 
to the CCP, though. The notion that objects without spatial extent must have 
(pointlike) forms arises if objects are treated as {\it phenomenal\/} objects, since 
formless objects cannot {\it appear\/}. And the notion that the ultimate 
constituents of matter are distinct substances is bolstered by the following facts: 
(i)~since phenomenal features that are present in the same place get integrated 
into the same phenomenal object, different phenomenal objects cannot be in the 
same place, (ii)~phenomenal objects are macroscopic, (iii)~macroscopic objects 
change in ways that ensure that they are re-identifiable.

Nor is it to be expected that anyone would try to incorporate all of the items in the 
right column in his or her model of the physical world, not least because some of 
them are mutually inconsistent. For instance, if the ultimate constituents of matter 
are pointlike, the shapes of things with spatial extent cannot be surfaces. Yet, to 
the best of my knowledge, the interpretation outlined in this chapter is the only one 
that rejects all of those items. Specifically, the postulate of an intrinsically and 
completely differentiated, pre-existent spatial or spatiotemporal expanse appears 
to be shared by all other models.

\section{\large SPACE, TIME, AND THE WAVE FUNCTION}
\label{stwf}
It strikes me as odd that the ontological and/or epistemological status of the wave 
function has been the focus of a lively controversy for three quarters of a century, 
while the ontological status of the coordinate points and instants on which 
$\psi(x,t)=\braket x{\psi(t)}$ functionally depends has hardly ever been called in 
question.

Consider the state $\ket{z_+}$. It is a shorthand notation for the projector 
$\ketbra{z_+}{z_+}$ representing (i)~a possible outcome of a measurement of 
the $z$~component of the spin of (say) an electron and (ii)~the density operator 
``prepared'' by this outcome. It informs us of the outcome of a measurement, and 
it allows us to compute the probabilities of the possible outcomes of subsequent 
measurements. Nothing can be said without reference to measurements. Why? 
Because an ``apparatus'' is needed to {\it realize\/} an axis. By realizing an axis, 
the apparatus makes available two possible values; it {\it creates\/} possibilities 
to which probabilities can be assigned. In its absence, the properties ``up'' and 
``down'' do not even exist as possibilities. The notion that the 
symbol~$\ket{z_+}$ represents something {\it as it is\/}, rather than as it 
behaves in any given measurement context, is wrong.

The same goes for~$\psi(x,t)$. As we saw in Sec.~\ref{impdet}, in the absence of 
a detector that realizes a region~$V$, or the distinction between ``inside~$V$'' 
and ``outside~$V$'', the properties of being inside~$V$ and being outside~$V$ do 
not even exist as possibilities.

Alas, the existence of an intrinsically and completely differentiated, pre-existent 
space or spacetime appears to be taken for granted by all ontologizing interpreters 
of QM. In view of the incomplete spatiotemporal differentiation of the physical 
world implied by the quantum formalism itself, it should therefore come as no 
surprise that the interpretation of QM is beset with pseudoquestions and gratuitous 
answers. As long as the existence of an intrinsically and completely differentiated 
background spacetime is assumed, it is safe to say that our attempts to beat sense 
into QM are doomed.

\section{\large THE PSYCHOLOGY OF QUANTUM STATE\\
EVOLUTION}
In addition to the neurobiological reasons why we are prone to think of space (and 
therefore of spacetime as well) as an intrinsically differentiated expanse, there are 
strong psychological reasons why we are prone to think of time as intrinsically 
differentiated.%
\footnote{We will see in a moment why these reasons cannot be neurobiological.}
In brief: while the successive nature of our experience makes it natural for us to 
hold that only the present is real, or that it is somehow ``more real'' than the 
future and the past, our self-experience as agents makes it natural for us to hold 
that the known or in principle knowable past is ``fixed and settled'', while the 
unknown future is ``open''.

Because it is impossible to consistently project the experiential Now into the 
physical world, none of this has anything to do with physics. We are accustomed to 
the idea that the redness of a ripe tomato exists in our minds, rather than in the 
physical world. We find it incomparably more difficult to accept the same as true of 
the experiential Now: it has no counterpart in the physical world. There simply is no 
objective way to characterize the present.%
\footnote{This is why the special experiential character of the present, which 
induces us to think of time as intrinsically differentiated, cannot be understood in 
neurobiological terms.}
The temporal modes past, present, and future can be characterized only by how 
they relate to us as conscious subjects: through memory, through the 
present-tense immediacy of sensory qualities, or through anticipation. In the 
physical world, we may qualify events or states of affairs as past, present, or future 
{\it relative to\/} other events or states of affairs, but we cannot speak of {\it 
the\/} past, {\it the\/} present, or {\it the\/} future. The idea that some things 
exist not yet and others exist no longer is as true (psychologically speaking) and as 
false (physically speaking) as the idea that a ripe tomato is red.%
\footnote{If we conceive of temporal or spatiotemporal relations, we conceive of 
the corresponding relata simultaneously---they exist at the same time {\it in our 
minds\/}---even though they happen or obtain at different times in the physical 
world. Since we cannot help it, that has to be OK. But it is definitely not OK if we 
sneak into our simultaneous mental picture of a spatiotemporal whole anything 
that advances across this spatiotemporal whole. We cannot mentally represent a 
spatiotemporal whole as a simultaneous spatial whole and then imagine this 
simultaneous spatial whole as persisting and the present as advancing through it. 
There is only one time, the fourth dimension of the spatiotemporal whole. There is 
not another time in which this spatiotemporal whole persists as a spatial whole and 
in which the present advances. If the experiential now is anywhere in the 
spatiotemporal whole, it is trivially and vacuously everywhere---or, rather, 
everywhen.}

In the physical world, time does not ``flow'' or ``pass''. To philosophers, the 
perplexities and absurdities entailed by the notion of a changing objective present 
or a flowing time are well known. (See, e.g., the illuminating entry on ``time'' in 
Ref.~\cite{Audi}.) To physicists, the subjectivity of a temporally unextended yet 
persistent and continually changing present was brought home by the relativity of 
simultaneity. The same is implied by QM, inasmuch as the incomplete 
differentiation of the quantum world rules out the existence of an evolving {\it 
instantaneous\/} physical state.

Again, the physical correlation laws (whether classical or quantum) know nothing 
of a preferred direction of causality. They are time-symmetric. They permit us to 
retrodict as well as to predict. The figment of a causal arrow is a projection, into 
the physical world, of our sense of agency, our ability to know the past, and our 
inability to know the future. It leads to the well-known folk tale according to which 
causal influences reach from the nonexistent past to the nonexistent future 
through persisting ``imprints'' on the present: If the past and the future are 
unreal, the past can influence the future only through the mediation of something 
that persists. Causal influences reach from the past into the future by being 
``carried through time'' by something that ``stays in the present''. This evolving 
instantaneous state includes not only all presently possessed properties but also 
traces of everything in the past that is causally relevant to the future.

In classical physics, this is how we come to conceive of fields of force that evolve in 
time (and therefore, in a relativistic world, according to the principle of local 
action), and that mediate between the past and the future (and therefore, in a 
relativistic world, between local causes and their distant effects). Classical 
electrodynamics is a case in point. While at bottom it is nothing but an algorithm 
for calculating the effects that electrically charged objects have on electrically 
charged objects, the projection of an evolving instantaneous state into the world of 
classical physics forces us to transmogrify this algorithm into a continuous, local, 
physical process by means of which effects are produced.

The projection of an evolving instantaneous state into the world of quantum 
physics forces us to seize instead on a probability algorithm that depends on the 
relative times between measurements and on the outcomes of earlier 
measurements, to transmogrify the same into an instantaneous physical state, and 
to think of its evolution as a physical process that plays a similar mediating role.

The all but universally accepted projection into the quantum world of the 
intrinsically differentiated time ``continuum'' of classical physics made it a 
foregone conclusion that $\psi$ would come to be thought of as an evolving, 
instantaneous physical state. Since von Neumann's influential work~\cite{vN}, 
textbooks list versions of the following claims among the axioms of QM:
\bi
\item[(X)]Between measurements, quantum states evolve according to unitary 
transformations (or according to trace-preserving completely positive linear maps) 
and thus continuously and predictably.
\item[(Y)]By way of measurement, quantum states evolve as stipulated by the 
projection postulate (or, up to normalization, via outcome-dependent completely 
positive linear maps) and thus in general discontinuously and unpredictably.
\ei
While the real trouble with these claims is that they postulate two modes of 
evolution rather than {\it none\/}, virtually everybody agrees that the trouble 
with (standard) QM is the postulation of two modes of evolution rather than {\it 
one\/}. Unitary evolution is considered ``normal'' and therefore not in need of 
explanation. Hence the mother of all pseudoproblems: how to explain (away) the 
unpredictable ``collapses'' of wave functions due to measurements?

Stripped of the notion that quantum states evolve, (X) and (Y) merely state the 
obvious. An algorithm for assigning probabilities to possible measurement 
outcomes on the basis of actual measurement outcomes has {\it two\/} perfectly 
normal dependences: It depends continuously on the times of measurements. (If 
you change the time of a measurement by a small amount, the probabilities 
assigned to the possible outcomes change by small amounts). And it depends 
discontinuously on the outcomes that constitute the assignment basis. (If you take 
into account an outcome that was not previously taken into account and had a prior 
probability less than~1, the assignment basis changes unpredictably as a matter of 
course, and so do the probabilities assigned.)

\section{\large THE WORLD BETWEEN MEASUREMENTS}
\label{wbm}
There is a notion that probabilities are inherently subjective, and that therefore QM 
{\it qua probability algorithm\/} is an epistemic theory concerned with knowledge 
or information~\cite{Petersen,Peierls}. It does not yield a model of a 
``free-standing'' reality~\cite{FuPer}. This is wrong. That probabilities are 
inherently subjective, is a wholly classical idea. The very fact that the fundamental 
theoretical framework of contemporary physics is a probability algorithm, signals 
that the probabilities it serves to assign are {\em objective\/}~\cite{piqm,ABL}. 
They are inevitable ingredients in any adequate description of the quantum world. 
Subjective probabilities are ignorance probabilities. They enter the picture when 
relevant facts are ignored. They disappear when all relevant facts are taken into 
account. The ``uncertainty'' principle guarantees that quantum-mechanical 
probabilities cannot be made to disappear. The reason this is so is not a lack of 
knowledge but a lack of relevant facts: the totality of earlier measurement 
outcomes is insufficient for predicting subsequent measurement outcomes with 
certainty. The stability of ordinary matter---the existence of objects that have 
spatial extent, are composed of a (large but) finite number of objects without 
spatial extent, and neither collapse nor explode as soon as they are created%
\footnote{If one looks for an appropriate formalism for quantifying the fuzziness 
that ``fluffs out'' matter, one is lead more or less directly to the probability 
algorithm of standard QM~\cite{JustSo}---another good reason why the quantum 
formalism is fundamentally and irreducibly a probability algorithm.}%
---hinges on the {\em objective fuzziness\/} of their internal relative positions 
and momenta, not on our subjective uncertainty about the values of these 
variables. (The literal meaning of Heisenberg's ``Unsch\"arfe'' {\em is\/} 
``fuzziness''.)

What is the proper (i.e., mathematically rigorous and philosophical sound) way of 
describing fuzzy variables? It is to assign objective probabilities (not to be 
confused with relative frequencies) to the possible outcomes of measurements. 
Consider again the spin ``state''~$\ket{z_+}$. Suppose that after the 
measurement that warrants the use of this algorithm, no further measurement is 
made. There is a notion that if QM concerns nothing but correlations between 
measurement outcomes, nothing can be said about the actual state of a spin (or 
any other system, for that matter) if no measurement is {\it actually\/} made. 
This, too, is wrong. While it is correct that nothing can be said without reference to 
measurements, a complete description of the fuzzy state of affairs that obtains 
after the measurement can be given; it consists in the probabilities that 
$\ket{z_+}$ assigns to the possible outcomes of {\it unperformed\/} 
measurements~\cite{clicks,iucaa}. $\ket{z_+}$~describes a fuzzy orientation by 
means of counterfactual probability assignments that vary continuously from~1 
(for ``up along the $z$~axis'') to~0 (for ``up along the inverted $z$~axis'') as 
the axis of measurement (defined by the gradient of $|{\bf B}|$) is rotated by 
$180^\circ$ about any axis perpendicular to the $z$~axis. At any rate, this is the 
fuzzy state of affairs that obtains after a measurement has yielded the outcome 
$z_+$ {\it if\/} the Hamiltonian is~0, and {\it if\/} no further measurement is 
subsequently made.

If the Hamiltonian is not~0, then the probability assignments describing the 
subsequent fuzzy state of affairs depend on the times of (unperformed) 
measurements. This is not the same as saying that the subsequent fuzzy state of 
affairs changes with time, for the antecedents of these counterfactual probability 
assignments are false not only because they affirm that a measurement is made, 
but also because they affirm that this is made {\it at a particular time\/}. Where 
the electron's spin is concerned, there is no particular time until another 
measurement is made. The quantum world is built from the top-down 
(Sec.~\ref{tds}), and temporal as well as spatial distinctions are relative and 
contingent (Sec.~\ref{rcr}). The electron's spin is temporally differentiated by the 
events that indicate its values. Between actual measurements it is only 
counterfactually differentiated (by unperformed measurements).

If another measurement is subsequently made, the fuzzy state of affairs that 
obtains in the meantime (during which no measurements are made) is fully 
described only if all relevant information is taken into account. This includes the 
outcome of the subsequent measurement. (Probability assignments based on 
earlier and later measurement data are made according to the ABL 
rule~\cite{piqm,ABL}.)

How can a measurement outcome contribute to determine a state of affairs that 
obtains during an {\it earlier\/} time? This ceases to be a mystery if the following 
points are taken into account. (i)~There is no instantaneous physical state that 
evolves and thereby introduces an arrow of time. (ii)~The relevant laws are 
time-symmetric. Born probabilities can be assigned on the basis of later as well as 
earlier outcomes, and ABL probabilities are by nature time-symmetric. 
Measurement outcomes can therefore be used to assign probabilities to the 
possible outcomes of unperformed measurements before, after, and between actual 
measurements. And these assignments can contribute to describe the fuzzy states 
of affairs that obtain before, after, and between actual measurements. As said, the 
notion that later states of affairs are determined only by earlier states of affairs is 
nothing but a psychological projection of our self-experience as agents in a 
successively experienced world.

An arrow of time does exist, but it pertains to the causal nexus in which 
measurements or value-indicating events (VIEs) are embedded (see 
Sec.~\ref{profac}). For inasmuch as VIEs create traces or records (which is 
necessary for their being value-{\it indicating\/} events) they are causally linked 
to the future, and there are two senses in which they are causally decoupled from 
the past. For one, the particular outcome of a successful measurement is generally 
not necessitated by any antecedent event. For another, nothing guarantees the 
success of an attempted measurement. Since every quantum-mechanical 
probability assignment {\it presupposes\/} the (actual or counterfactual) 
occurrence of a~VIE, QM~cannot supply sufficient conditions for the occurrence of 
a~VIE.%
\footnote{While this implies that perfect detectors are a fiction, it does not prevent 
us from invoking perfect detectors to assign probabilities to the possible outcomes 
of {\it unperformed\/} measurements.}
VIEs~are {\it uncaused\/}.%
\footnote{But isn't the click caused by the ionization of an atom in the counter? 
No, for while it is true that without an ionization there would be no click, it is 
equally true that without a click there would be no 
ionization~\cite{clicks,UlfBohr}. The microworld supervenes on the macroworld.}

The bottom line: What obtains between measurements is what is appropriately 
described by counterfactual probability assignments, namely, fuzzy states of 
affairs. Like the probability measure that describes it, a fuzzy state of affairs is not 
something that evolves. It is not only fuzzy but also temporally undifferentiated.

\section{\large THE PROBLEM OF FACTUALITY}
\label{profac}
A fundamental physical theory that is essentially an algorithm for assigning 
probabilities to VIEs on the basis of other VIEs presupposes the occurrence of VIEs, 
and the challenge is to demonstrate that such a theory can be complete, in the 
sense that it not only presupposes but also encompasses the VIEs. The challenge is 
{\it not\/} to explain how possibilities---or worse, 
probabilities~\cite{Treiman}---become facts, how properties 
emerge~\cite{JoosZeh}, or why events occur~\cite{Pearle}. Saying in common 
language that a possibility becomes a fact is the same as saying that something 
that is possible---something that {\it can\/} be a fact---actually {\it is\/} a fact. 
This non-problem becomes a pseudoproblem if the common-language ``existence'' 
of a possibility is construed as a lesser kind of existence---a matrix of 
``propensities''~\cite{Popper} or ``potentialities''~\cite{Heisenberg, Shimony} 
that transform into the genuine article (nonexistence proper or existence proper) 
by way of measurement. It is the sort of problem that is bound to arise if one 
misconstrues a probability algorithm as an evolving physical state. Our task as 
measurement theorists is not to account for the occurrence of VIEs,%
\footnote{Because VIEs are uncaused, this is as impossible as explaining why 
there is anything at all, rather than nothing.}
let alone for the actualization of possibilities independently of VIEs, but to identify 
that substructure of the theoretical structure of QM which encompasses the VIEs, 
and which can be consistently considered factual {\it per se\/}.

Here goes: The possibility of obtaining evidence of the departure of an object~$O$ 
from its classically predictable position, given all relevant earlier 
position-indicating events, calls for detectors whose position probability 
distributions are narrower than~$O$'s. Such detectors do not exist for all objects. 
Some objects have the sharpest positions in existence. For these objects, the 
probability of obtaining such evidence is extremely low. Hence {\em among\/} 
these objects there are many of which the following is true: every one of their 
indicated positions is consistent with (i)~every prediction that is based on their 
previous indicated positions and (ii)~a classical law of motion. Such objects 
deserve to be called ``macroscopic''. To enable a macroscopic object to indicate an 
unpredictable value, one exception has to be made: its position may change 
unpredictably if and when it serves to indicate such a value.

Decoherence investigations~\cite{Zurek2003, Blanchard2000, Joos2003} have 
demonstrated for various reasonable definitions of ``macroscopic'' that the 
probability of finding a macroscopic object where classically it could not be, is 
extremely low. This guarantees the abundant existence of macroscopic objects 
according to the present, stricter definition, which are never actually ``found'' 
where classically they could not be. The correlations between the indicated 
positions of these objects are deterministic in the following sense: their fuzziness 
never evinces itself through outcomes that are inconsistent with predictions that 
are based on earlier outcomes and a classical law of motion. Macroscopic objects 
(including pointers) follow trajectories that are only counterfactually fuzzy. That is, 
they are fuzzy only in relation to an imaginary background that is more 
differentiated spacewise than is the actual physical world. In the latter, there are 
no regions over which they are ``smeared out''. So we cannot say that they are 
fuzzy---nor can we say that they are sharp: ``not fuzzy'' implies ``sharp'' only if 
we postulate the intrinsically and completely differentiated background space that 
the quantum world lacks.

To be able to identify a substructure of the theoretical structure of QM that can be 
consistently considered factual {\it per se\/}, we must be allowed to look upon 
the positions of macroscopic objects---macroscopic positions, for short---as 
intrinsic, as self-indicating, or as real {\it per se\/}. The reason why this is both 
possible and legitimate, is that the extrinsic nature of the values of physical 
variables is implied by their fuzziness. If macroscopic positions aren't fuzzy, we 
have every right to consider them intrinsic---notwithstanding that they are {\it 
also\/} extrinsic, for even the Moon has a position only because of the myriad of 
``pointer positions'' that betoken its whereabouts. The reason why macroscopic 
positions can be both extrinsic and intrinsic is that they indicate each other so 
abundantly, so persistently, and so sharply that they are only counterfactually 
fuzzy.

While the extrinsic nature of macroscopic positions forbids us to think of an {\it 
individual\/} macroscopic position as intrinsic or real {\it per se\/}, nothing 
stands in the way of attributing to the {\it entire\/} system of macroscopic 
positions an {\it independent\/} reality, and nothing prevents us from considering 
the {\em entire\/} system of existing spatial relations or possessed relative 
positions (including the corresponding multitude of relata) as 
{\it self-contained\/}. What the extrinsic nature of the properties of the quantum 
world forbids is to model this world from the bottom up. The macroscopic does not 
supervene on the microscopic, whether conceived as a multitude of substances or 
as a transfinite manifold of spacetime points. The ``foundation'' is above. The 
properties of the microworld exist because they are indicated by the goings-on in 
the macroworld. They supervene on the macroworld---the system of macroscopic 
positions, in which value-indicating events occur as unpredictable transitions, in 
compliance with the quantum-mechanical correlation laws.

The system of macroscopic positions thus is a substructure in two senses: it is a 
part of the entire theoretical structure of QM, and it is the self-existent foundation 
on which all indicated values supervene. What makes the macroworld the sole 
candidate for the predicate ``factual {\it per se\/}'' is the physical unreality of its 
own fuzziness or (equivalently) the fact that the spatial differentiation of the 
physical world is incomplete. Even though there is no hermitian ``factuality 
operator'' (factuality cannot be measured), QM thus uniquely determines what is 
factual {\it per se\/}. Owing to the crucial role played the incomplete spatial 
differentiation of the physical world, however, this cannot be seen, and therefore 
the notorious measurement problem cannot be solved, as long as a completely 
differentiated spatial background is postulated.

\section{\large FUZZINESS: IN THE MIND OR IN THE WORLD?}
In discussions of the measurement problem, the following unitary ``transition'' 
usually plays as central role:
\be
\sum_i c_i\ket{q_i}\otimes\ket{A_0}\longrightarrow\sum_i 
c_i\ket{q_i}\otimes\ket{A_i}.
\label{e1}
\ee
$A_i$ is the property whose possession, by an apparatus~$\cal A$, indicates that a 
system~$\cal S$ has the property (or an observable~$Q$ has the value)~$q_i$, 
and $A_0$ is the apparatus-property of being in the neutral state. This substitution 
of one probability measure for another takes care of the time between a pair of 
measurements informing us how $\cal A$ and~$\cal S$ are ``prepared'', and 
another pair of measurements performed on $\cal A$ and~$\cal S$, respectively, 
without taking into account the outcomes of these later measurements. Because of 
the perfect correlations between their possible outcomes, we are entitled to 
interpret the possession (by the apparatus) of a particular~$A_i$ as indicating the 
possession (by the system) of the corresponding~$q_i$, and because the 
apparatus-properties $A_i$ involve macroscopic positions, we are entitled to 
regard one (and only one) $A_i$ as possessed.

If the substitution (\ref{e1}) is misconstrued as a physical transition from the 
initial state of affairs $\sum_i c_i\ket{q_i}$ to the final state of affairs $\sum_i 
c_i\ket{q_i}\otimes\ket{A_i}$, the following pseudoquestions arise: How is it 
that measurements appear to have outcomes? And why does the outcome 
correspond to an element of this particular decomposition of the final state?

The decomposition of the final algorithm (\ref{e1}) is biorthogonal: the kets 
$\ket{q_i}$ are mutually orthogonal because $Q$ is hermitian, and the kets 
$\ket{A_i}$ are mutually orthogonal because $\cal A$'s possession of $A_i$ 
indicates not only the truth of ``$Q$~has the value~$q_i$'' but also the falsity of 
``$Q$~has the value~$q_k$'' for $k\neq i$. Modal interpretations 
(MIs)~\cite{VF,Dieks3,Vermaas} capitalize on the uniqueness of the biorthogonal 
decomposition in the event that all $c_i$ have different norms. To ensure that one 
element of the decomposition represents reality, MIs simply postulate that 
whenever a two-component system has a unique biorthogonal decomposition, 
exactly one term of this decomposition represents the actual state of the system.

Spontaneous localization theories (SLTs)~\cite{GRW,Pearle1,Pearle2} introduce 
nonlinear or stochastic modifications of the standard dependence of probabilities 
on the times of measurements, in order to ensure that the probabilities associated 
with value-indicating pointer positions are either close to~1 or close 
to~0---needless to say, without enabling us to predict which pointer position will 
be the lucky one; betting on outcomes remains guided by the Born probabilities of 
standard~QM.

Decoherence theories (DTs)~\cite{JoosZeh,Zurek} capitalize on the fact that all 
known interaction Hamiltonians contain $\delta$-functions of the distances 
between particles. This makes the environment a more accurate monitor of pointer 
positions than any individual apparatus could be. Individual measurements of 
pointer positions therefore reveal pre-existent properties, in the sense that they 
indicate properties that are monitored by the environment.%
\footnote{In the vocabulary of decoherence theorists, ``monitored by the 
environment'' is synonymous with something like ``decoherently correlated with 
the environment''.}

One common characteristic of these interpretative strategies is that an ever so 
small quantitative difference is held to be sufficient for a considerable conceptual 
difference. Consider a biorthogonal probability measure.
\be
a_1\ket{\phi_+}\otimes\ket{\psi_+}+a_2\ket{\phi_-}\otimes\ket{\psi_-}
\label{e2}
\ee
If the norms of the coefficients $a_1$ and~$a_2$ differ ever so little, then, 
according to MIs, one element of this decomposition represents the actual state. 
Otherwise they don't~\cite{Dieks3}. An ever so slight difference thus can decide 
whether or not a value exists. SLTs (like many other approaches) subscribe to the 
notion that probability~1 is necessary and sufficient for ``has'' or ``is''. This 
means that the radical difference between either ``has'' or ``lacks'' and ``neither 
has nor lacks'' can hinge on an ever so slight difference between a probability equal 
to~1 and a probability less than~1. DTs are motivated by the belief that the 
vanishing of off-diagonal terms is necessary and sufficient for the reinterpretation 
of a density operator as a proper mixture. An ever so small difference between $=$ 
and $\approx$ can therefore make the enormous conceptual difference between 
``and'' and ``or''.%

Another unpalatable feature of the interpretative strategies considered here arises 
as a consequence. These strategies countenance a {\it conceptual\/} fuzziness 
that permits $\approx$ to do duty for~$=$. The probabilities of VIEs in SLTs are 
never exactly~1 (the relevant probability distributions have ``tails''). DTs succeed 
in demonstrating that the relevant off-diagonal terms remain very small for very 
long times but not that they remain~0 for all time to come. MIs need to 
demonstrate that the final algorithm in (\ref{e1}) is biorthogonal. But this is 
tantamount to demonstrating that the off-diagonal terms vanish, as one gathers 
from the final density operator associated with~$\cal S$,
\be
\sum_{ij}c_ic_j^*\braket{A_j}{A_i}\ketbra{q_i}{q_j},
\ee
which is diagonal (regardless of the values of the~$c_i$) if and only if 
$\braket{A_j}{A_i}=0$ for $i\neq j$.

This conceptual fuzziness, as d'Espagnat has argued at length and 
convincingly~\cite{dE1,dE2,dE3}, is unacceptable in a strongly objective theory. 
Therefore, so d'Espagnat, QM is an objective theory only in a weak, intersubjective 
sense of ``objective''. Similar qualifications have been made by many other 
authors. While Zurek advocates a twofold relativization of 
``existence''~\cite{Zurek2003,Zurek,MohrhoffTEOE}, Zeh~\cite{Zeh} argues that 
``[w]hile decoherence transforms the formal `plus' of a superposition into an 
effective `and' (an apparent ensemble of new wave functions), this `and' becomes 
an `or' only with respect to a subjective observer''. Such qualifications become 
necessary only if the world's incomplete spatiotemporal differentiation (implied by 
the fuzziness of all relative positions) is ignored and a completely differentiated 
background spacetime is postulated instead. The choice is between fuzziness in the 
head or fuzziness in the world.

\section{\large THE LIMITS OF EXPLANATION}
Can we hope to explain the quantum-mechanical correlation laws? If these laws are 
indeed the fundamental laws of physics (and apart from our dogged insistence on 
explaining from the bottom up, we have no reason to believe that they are not), 
then they cannot be explained the way Kepler's laws of planetary motion can be 
explained by Newton's law of gravity. Only a law that is not fundamental can be so 
explained. What is more, QM does not permits us to decompose the quantum world 
into a multitude of interacting substances with causally connected properties. The 
explanatory framework provided by these concepts, so useful for our dealings with 
the phenomenal world, cannot be expanded to include more than the macroworld. 
To look for causal explanations of the quantum-mechanical correlations is to put 
the cart before the horse. It is the fundamental correlation laws that define the 
extent to which causal concepts may be used.

Can we interpret the quantum-mechanical correlation laws as {\em descriptive\/} 
of a mechanism or physical process? As far as the {\it synchronic\/} correlations 
(``EPR correlations'') are concerned, this possibility seems too remote for serious 
consideration. One only has to consider the state of three spin-$1/2$ particles 
discussed Greenberger, Horne, and Zeilinger~\cite{GHZ}. If this state is prepared, 
it is possible to predict {\it any\/} spin component of {\it any\/} particle by 
subjecting the two other particles to appropriate measurements, even though the 
correlations between the possible outcomes of spin measurements are independent 
of the distances between the three particles, and even though it is impossible to 
think of these measurements as revealing pre-existent values. It is hoped that this 
chapter has made it clear why attempts at interpreting the {\it diachronic\/} 
correlation laws as descriptive of some physical process, are equally misguided.

There remains the question of how the correlations are possible at all. We tend to 
believe, as Einstein did, that ``things claim an existence independent of one 
another'' whenever they ``lie in different parts of space''~\cite{Einstein}. If this 
were true, the correlations would indeed be impossible. Fact is that the three 
particles, irrespective of the distances between them, are {\em not\/} 
independent of one another. Fiction is that they lie in different parts of space. In 
the quantum world, space has no parts, whether we think of space as a set of 
relations (a set of relations has no parts) or as a pre-existent expanse (QM does 
not permit us to think of it as divided). The quantum world, moreover, has room for 
only one substance. Considered by themselves, the ultimate constituents of matter 
are identical in the strong sense of numerical identity. All existing relations are 
self-relations. How, then, could things possibly ``claim an existence independent 
of one another''?

Einstein based his belief in the mutual independence of objects situated in different 
parts of space on the demand that these objects be independent of the perceiving 
subject~\cite{Einstein}. This is ironic, for it is precisely the illegitimate projection 
of the structure of the phenomenal world into the physical world that underlies this 
belief. In the phenomenal world, which is constructed in conformity with the CCP, 
places pre-exist (Sec.~\ref{ttw}), and features that are present in different places 
get integrated into different phenomenal objects (which may be parts of the same 
object). This is why we are convinced by default that whatever exists in different 
places must be different objects. If the perceived separateness of things situated in 
different parts of space is given the status of a fundamental ontological truth, 
things can influence each other only by some kind of direct contact, across common 
boundaries. It is to this naive idea that we give the grand name ``principle of local 
action''.%
\footnote{``[P]hysicists are, at bottom, a naive breed, forever trying to come to 
terms with the `world out there' by methods which, however imaginative and 
refined, involve in essence the same element of contact as a well-placed 
kick''~\cite{DeWGra}.}

\section{\large CONCLUSION}
At present, the physics community can be divided into three factions. The first---the 
majority---doesn't care very much what (if anything) QM is trying to tell us about 
the nature of Nature. The second resigns itself to agnosticism. It asserts that we 
cannot describe the quantum world as it is (by itself): its features are forever 
beyond our ken. All we can usefully talk about is statistical correlations between 
measurement outcomes. The third aspires to describe the quantum world as it is, 
independent of measurements. This faction is split into numerous sects, each 
declaring to see the light, the ultimate light. Go to any conference on quantum 
foundations, and you will find their priests pitted in holy war~\cite{Fuchs2002}.

The agnostics and the priests both have a point and both are wrong. The agnostics 
have a point in that nothing of relevance can be said without reference to 
measurements. They are wrong in claiming that the features of the quantum world 
are beyond our ken. The priests have a point in that it is indeed possible to describe 
the features of the quantum world. They are wrong in claiming that these features 
can be described without reference to measurements.

We have examined some of the neurobiological and psychological reasons for this 
sorry state of affairs. Because of them, it is virtually impossible to recognize the 
ontological implications of QM or (if they are recognized) to accept them. To many 
they might seem as preposterous as the metaphysical claims of the quantum-state 
realists. We shall therefore continue to hear claims to the effect that ``quantum 
theory does {\it not\/} describe physical reality. What it does is provide an 
algorithm for computing {\it probabilities\/} for the macroscopic events 
(``detector clicks'') that are the consequences of our experimental 
interventions''~\cite[original emphasis]{FuPer}. And we shall be told that it is 
impossible to distill from QM a model of a free-standing reality independent of our 
interventions.

Yet agnosticism is nothing but a cop-out. Science is driven by the desire to know 
how things {\it really\/} are. It owes its immense success in large measure to its 
powerful ``sustaining myth''~\cite{Mermin}---the belief that this can be 
discovered. Neither the ultraviolet catastrophe nor the spectacular failure of 
Rutherford's model of the atom made physicists question their faith in what they 
can achieve. Instead, Planck and Bohr went on to discover the quantization of 
energy and angular momentum, respectively. If today we seem to have reason to 
question our ``sustaining myth'', it ought to be taken as a sign that we are once 
again making the wrong assumptions, and it ought to spur us on to ferret them out. 
We may yet have to learn how to reconcile a free-standing reality with the 
supervenience of the ``quantum domain'' on the macroworld, but how could 
anyone know that it is impossible? It is, in fact, quite possible, provided we cease 
to obtrude upon the quantum world the intrinsically and completely differentiated 
spatiotemporal background of classical physics.

\end{document}